\documentclass[twocolumn,secnumarabic,amssymb, nobibnotes, aps, pra]{revtex4-1}

\usepackage{scalerel}
\usepackage[percent]{overpic}
\usepackage{yfonts}
\usepackage[normalem]{ulem}
\usepackage{mathptmx}
\usepackage{braket}
\usepackage{anyfontsize}
\usepackage{times,dsfont,amssymb,amsmath,amsthm,amsfonts,amsbsy,mathrsfs,bm,bbm,graphicx,graphics,epsfig,multirow,mathtools,color,bbold,empheq,enumerate}
\usepackage{xcolor,soul}
\usepackage{textcomp}

\newcommand*\zeroOrder{\includegraphics[height=6pt]{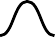}}
\newcommand*\firstOrder{\includegraphics[height=6pt]{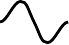}}

\begin{document}

\title{Remotely projecting states of photonic temporal modes}

\author{Vahid Ansari$^{1,2}$}
\email{vahid.ansari@uni-paderborn.de}
\author{John M. Donohue$^{1,3}$}
\author{Benjamin Brecht$^{1}$}
\author{Christine Silberhorn$^{1}$}

\affiliation{$^1$Integrated Quantum Optics, Paderborn University, Warburger Strasse 100, 33098 Paderborn, Germany}
\affiliation{$^2$E. L. Ginzton Laboratory, Stanford University, 348 Via Pueblo Mall, Stanford, California 94305, USA}
\affiliation{$^3$Institute for Quantum Computing, University of Waterloo, 200 University Ave W, Waterloo ON, N2L 3G1}

\begin{abstract}
Two-photon time-frequency entanglement is a valuable resource in quantum information. Resolving the wavepacket of ultrashort pulsed single-photons, however, is a challenge. Here, we demonstrate remote spectral shaping of single photon states and probe the coherence properties of two-photon quantum correlations in the time-frequency domain, using engineered parametric down-conversion (PDC) and a quantum pulse gate (QPG) in nonlinear waveguides. Through tailoring the joint spectral amplitude function of our PDC source we control the temporal mode structure between the generated photon pairs and show remote state-projections over a range of time-frequency mode superpositions.
\end{abstract}
\maketitle

\section{Introduction}

Quantum correlations provided by parametric down-conversion (PDC) photon pair sources are a powerful tool for quantum information science. The polarization, spatial, and time-frequency degrees of freedom can be employed to generate strong and verifiable two-photon entanglement~\cite{marcikic02timebin,schwarz2014experimental,kues2017chip,maclean2018direct}. These correlations enable techniques such as quantum state teleportation~\cite{boschi1998experimental,valivarthi2016quantum}, device-independent quantum key distribution~\cite{acin2007device}, and remote state preparation~\cite{bennett2001remote,peters2005remote,zavatta2006remote,baek2008temporal,koprulu2011lossless,averchenko2017temporal}. In order to exploit these resources for such tasks, it is necessary to have control over the generation of quantum correlations and also develop coherent measurement techniques in the desired degree of freedom.

While photonics provides the undisputed platform for implementations of multi-party quantum communication protocols and long-distance quantum experiments~\cite{shalm2015strong,zhong2015photon,ren2017ground}, each photonic degree of freedom carries associated advantages and challenges. The time-frequency degree of freedom, in particular, provides a high-dimensional quantum alphabet and is perfectly suited to fiber-based communication networks and integrated waveguide devices~\cite{Brecht2015,zhong2015photon,kues2017chip}. Entanglement in this degree of freedom is also naturally present in PDC sources, and can be controlled using pulse shaping techniques and material dispersion engineering~\cite{ansari2018tailoring}. However, the underlying time-frequency modes of the PDC state, also known as temporal Schmidt modes \cite{Law2000}, are not directly resolvable with traditional time or frequency measurements. Recently developed methods to control and manipulate the temporal mode structure of entangled states provide a powerful resource for entanglement-enabled photonic technologies~\cite{Eckstein2011,Brecht2014,Manurkar2016,ansari2017temporal,reddy2017temporal,ansari2018tomography}. However application of these methods to quantum states remains largely unexplored. 

\begin{figure}[t]
\centering
\includegraphics[width=.9\linewidth]{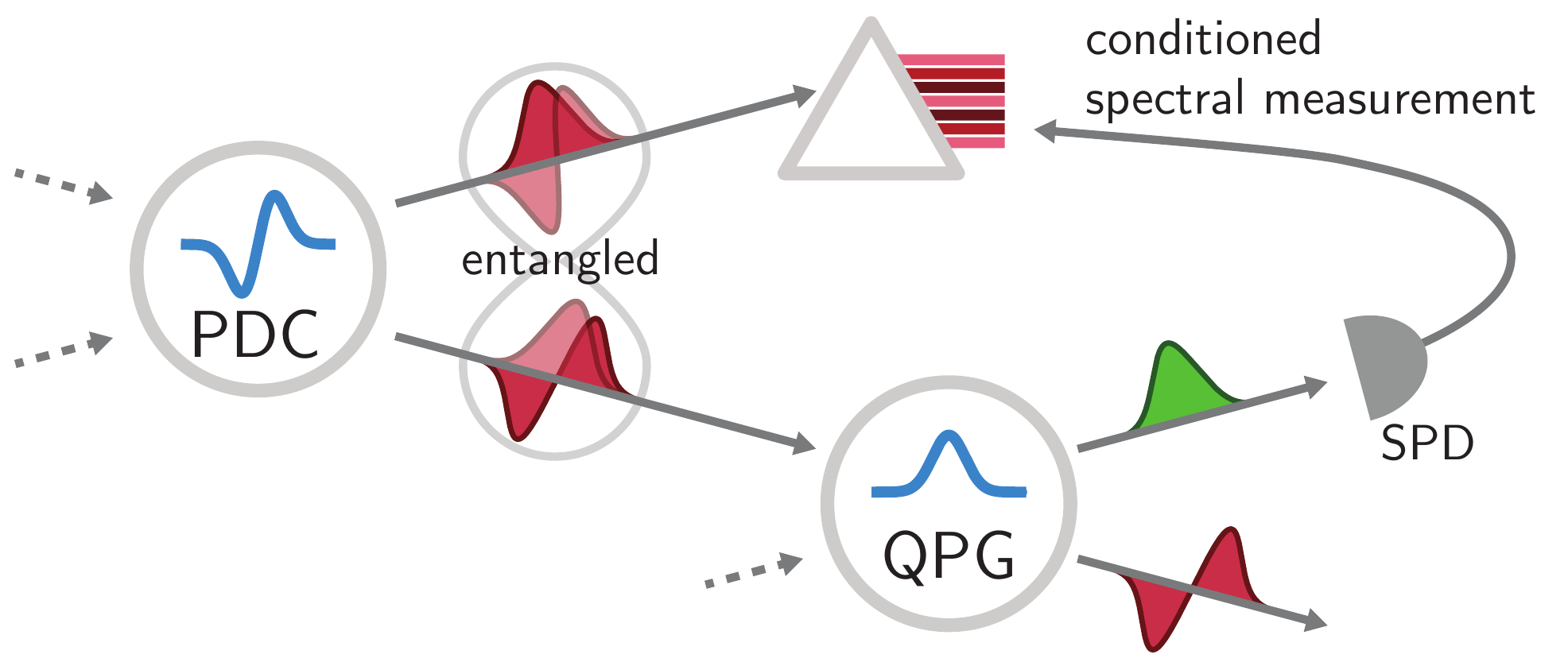}
\caption{Experimental concept. We generate time-frequency entangled photon pairs via parametric down-conversion (PDC), where the shape of the PDC pump pulse and the dispersion of the down-conversion medium allow us to engineer the exact form of the generated quantum correlations. We project upon chosen ultrashort pulsed time-frequency modes using a quantum pulse gate (QPG), which registers a successful projection by converting the signal to a visible (green) pulse. We then measure the spectrum of the partner photon conditioned upon detecting a green pulse on a single-photon detector (SPD).}
\label{fig:outline}
\end{figure}

In this work, we use tailored bipartite time-frequency quantum correlations to remotely prepare photonic temporal-mode states. Using a flexible toolbox of dispersion-engineered nonlinear optics and ultrafast pulse shaping, we perform projective measurements onto custom temporal-modes for one half of an entangled photon pair and measure the conditional spectrum of its partner, as sketched in Fig.~\ref{fig:outline}. We experimentally explore the correlated temporal-mode structure of PDC states with both traditional time-frequency correlations and engineered pulsed temporal-mode Bell-like correlations. In doing so, we also demonstrate that time-frequency remote state preparation can be used to efficiently prepare complex single-photon wave packets by harnessing the joint time-frequency coherence of the two-photon state.

\section{Engineering time-frequency entanglement}

\begin{figure*}[t]
\centering
\includegraphics[width=.8\linewidth]{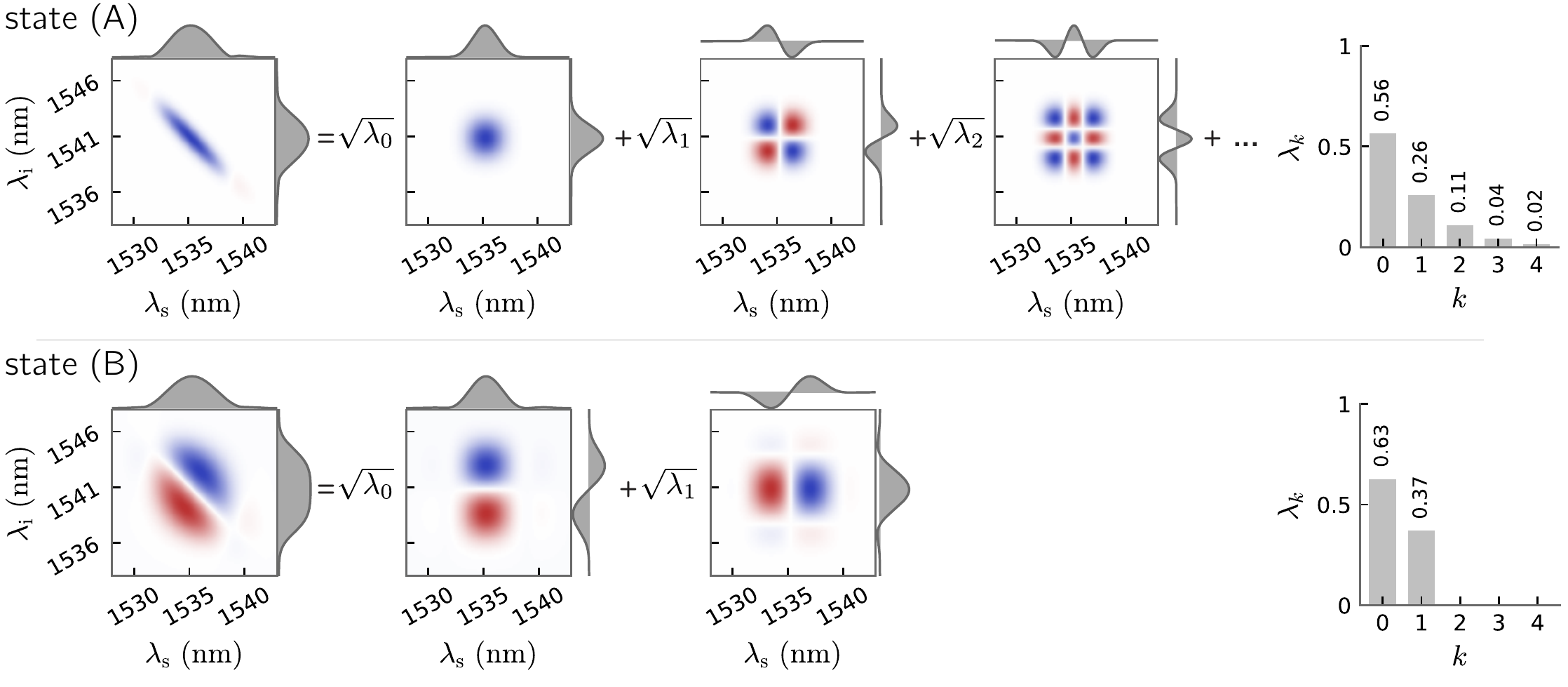}
\caption{The Schmidt decomposition of two-photon time-frequency correlations, modeled with parameters used in the experiment. (State A) When the PDC pump is narrow relative to the phasematching the Schmidt decomposition can be neatly approximated as a decaying series of matched Hermite-Gauss pulsed temporal-modes. (State B) By matching the pump and phasematching bandwidths and spectrally shaping the pump pulse, the number of generated Schmidt modes can be controlled, producing two-photon correlations similar to a time-frequency Bell state.}
\label{fig:schmidt}
\end{figure*}

We consider type-II PDC processes confined in waveguides with a single spatial mode, where we are interested in a subspace of the PDC state that has exactly one generated photon pair at any given time (and not multiple pairs of photons, nor the vacuum); these conditions can be experimentally matched by operating at low optical gains and performing coincidence measurements \cite{Eckstein2011b,Harder2013}. Then we can write the generated entangled state as 
\begin{equation}
|\psi\rangle_\mathrm{PDC}= \int d\omega_\mathrm{s}\, d\omega_\mathrm{i}\,f( \omega_\mathrm{s}, \omega_\mathrm{i})\hat{a}^\dagger(\omega_\mathrm{s})\hat{b}^\dagger(\omega_\mathrm{i})|0,0\rangle,
\label{eq:pdc}
\end{equation} where $\hat{a}^\dagger(\omega_\mathrm{s})$ and $\hat{b}^\dagger(\omega_\mathrm{i})$ define the signal (s) and idler (i) polarization modes and $f( \omega_\mathrm{s}, \omega_\mathrm{i})$ is the complex-valued joint spectral amplitude (JSA), normalized to ${\int d\omega_\mathrm{s}\,d\omega_\mathrm{i}\,|f(\omega_\mathrm{s},\omega_\mathrm{i})|^2=1}$. If the JSA is not factorizable, i.e. ${f( \omega_\mathrm{s}, \omega_\mathrm{i})}\neq{f_s( \omega_\mathrm{s})f_i(\omega_\mathrm{i})}$, the signal and idler share time-frequency entanglement. We can depict this entanglement more explicitly by considering the Schmidt decomposition of the state, equivalent to the singular value decomposition of the JSA. To do so, we decompose the JSA into a sum of orthonormal functions $\left\{g_k(\omega_\mathrm{s})\right\}$ and $\left\{h_k(\omega_\mathrm{i})\right\}$~\cite{Law2000,ansari2018tailoring}, 
\begin{equation}
f( \omega_\mathrm{s}, \omega_\mathrm{i} ) = \sum_k \sqrt{\lambda_k} g_k(\omega_\mathrm{s})h_k(\omega_\mathrm{i}).
\label{eq:pdc_schmidt}
\end{equation}
Eq.~\eqref{eq:pdc_schmidt} describes correlations such that if the signal photon is measured in the pulsed temporal-mode defined by the broadband field amplitude $g_k(\omega_\mathrm{s})$, the idler will collapse to the corresponding field amplitude $h_k(\omega_\mathrm{i})$.
The shape and decomposition of the JSA depends on the spectral shape of the PDC pump pulse, $\alpha(\omega_\mathrm{s}+\omega_\mathrm{i})$, and the phasematching function of the down-conversion medium, $\Phi(\omega_\mathrm{s},\omega_\mathrm{i})$, as 
\begin{equation}
f( \omega_\mathrm{s}, \omega_\mathrm{i} )=\alpha(\omega_\mathrm{s}+\omega_\mathrm{i})\Phi(\omega_\mathrm{s},\omega_\mathrm{i}).
\label{eq:jsa}
\end{equation} 
For traditional PDC states, the JSA exhibits frequency anti-correlations which are dominated by the energy conservation of the pump contribution. As shown in Fig.~\ref{eq:pdc_schmidt}(A), such a state is composed of many pulsed temporal-modes~\cite{Law2000}. If the JSA can be approximated by a two-dimensional Gauss function, the Schmidt decomposition gives correlated pairs of Hermite-Gauss temporal modes.

More interestingly, if the group velocities ($\mathrm{v}$) of the pump, signal, and idler are balanced such that $\frac{1}{\mathrm{v}_p}-\frac{1}{\mathrm{v}_s}=\frac{1}{\mathrm{v}_i}-\frac{1}{\mathrm{v}_p}$, it is possible to generate photon pairs in precisely one temporal-mode~\cite{Kuzucu2005,Harder2013}. This configuration, known as symmetric group velocity matching or extended phasematching, also allows for the generation of photon pairs with controlled entanglement dimensionality by only shaping the pump pulse~\cite{patera2012quantum,Brecht2015,ansari2018tailoring}. In particular, by shaping the pump pulse into the first order Hermite-Gauss shape with the same bandwidth as the phasematching function, the two-photon state emitted by the PDC source will be maximally-entangled with pulsed temporal Hermite-Gauss modes, analogous to the $\ket{\Psi^+}$ Bell state \cite{Brecht2015},
\begin{equation}
\ket{\Psi^+} = \frac{1}{\sqrt{2}}\left(\ket{\zeroOrder_\mathrm{s},\firstOrder_\mathrm{i}}+ \ket{\firstOrder_\mathrm{s},\zeroOrder_\mathrm{i}}\right).
\label{eq:state}
\end{equation}
In our experiment, the symmetric group-velocity matching is not perfectly matched which results in slightly different probabilities for the two terms, as seen in Fig.~\ref{eq:pdc_schmidt}(B). We also note that similar entangled states can be generated with shaping the crystal's non-linearity instead of the pump pulse \cite{graffitti_direct_2020}.

\section{Time-frequency mode measurements}

From the Schmidt decomposition of Eq.~\eqref{eq:pdc_schmidt} we can learn that if we condition an idler detection on a projection of the signal photon into the pulsed temporal-mode defined by $g_k(\omega_\mathrm{s})$, the idler will be found in the mode defined by $h_k(\omega_\mathrm{i})$. Considering the two-dimensional maximally entangled state of Eq.~\ref{eq:state}, if we project the idler photon into any arbitrary superposition ${\cos\theta\ket{\zeroOrder_\mathrm{i}}+\sin\theta\ket{\firstOrder_\mathrm{i}}}$, we will find the signal photon in the superposition state ${\sin\theta\ket{\zeroOrder_\mathrm{s}}+\cos\theta\ket{\firstOrder_\mathrm{s}}}$. This is known as remote state preparation~\cite{bennett2001remote,peters2005remote}, and can be extended to more complex remote shaping of time-frequency waveforms given a higher-dimensional entangled resource~\cite{sych2017generic,averchenko2017temporal,averchenko_efficient_2020}. Another interesting example is a highly multi-mode state, e.g. traditional frequency anti-correlated PDCs with very large Schmidt numbers. Equipped with appropriate methods, one can remotely reshape such PDC photons into arbitrary pulse shapes. By projecting one photon into a chosen composition of pulsed temporal-modes, we can herald the second photon in a target pulse shape. Existing methods to accomplish this rely on fast temporal modulation followed by frequency-resolved intensity detection~\cite{sensarn2009observation, averchenko2017temporal}, or are limited to time-bin encoded photons~\cite{zavatta2006remote}, or are based on intensity filtering~\cite{sedziak-kacprowicz_remote_2019}. An analogous technique has also been deployed to remotely herald single photons in particular spatial modes \cite{koprulu2011lossless}.

In our approach, the remote state preparation is achieved by coherent selection and measurement of pulsed temporal-modes, instead of intensity detection or filtering. This would also allow a high efficiency quantum pulse shaping because the remotely shaped single-photon wave packets are not subject to any physical modulation, which typically introduce loss and thus state degradation. Moreover, our technique allows coherent shaping with any time-frequency modal decomposition.

To directly project upon programmable temporal-modes, we use a sum-frequency generation between the idler photon and a shaped pump pulse in a dispersion-engineered waveguide, also known as a quantum pulse gate (QPG)~\cite{Eckstein2011,ansari2018tomography}. Dispersion engineering ensures that the input signal and pump are group-velocity matched but walk off significantly from the sum-frequency signal. Shaping the pump spectrum to $\gamma(\omega_\mathrm{p})$, the probability of a successful up-conversion is proportional to 
\begin{equation}
P_{\mathrm{up}}\propto\int\,d\omega_{\mathrm{s}}\gamma(-\omega_\mathrm{s})f_s(\omega_\mathrm{s}),
\label{eq:proj}
\end{equation} 
effectively implementing a projective measurement of the signal amplitude $f_s(\omega_\mathrm{s})$ onto the broadband time-frequency amplitude $\gamma^*(-\omega_\mathrm{s})$~\cite{ansari2018tailoring}. This technique is applicable to arbitrary superpositions of field-overlapping temporal modes and can directly access the temporal Schmidt modes of PDC photon pairs~\cite{ansari2017temporal}. In previous works, we have thoroughly shown coherence properties of the QPG and its high fidelity in implementing projective measurements into arbitrary time-frequency modes \cite{ansari2018tomography, Allgaier2017}.

\section{Experiment}

\begin{figure}[t]
\centering
\includegraphics[width=1\linewidth]{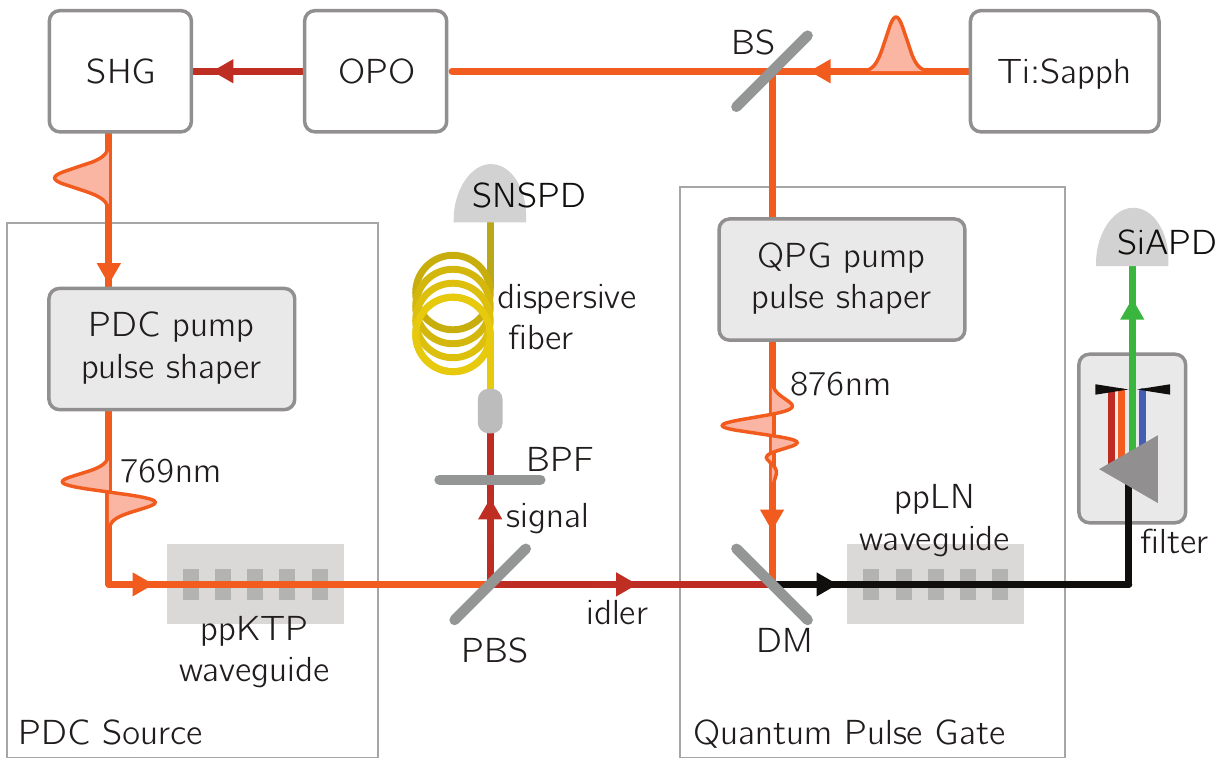}
\caption{Experimental setup. Pump pulses for the PDC and QPG process are shaped into Hermite-Gauss modes using SLM-based pulse shapers. Signal and idler photons are generated in a ppKTP waveguide, and the signal photon is measured using a quantum pulse gate (QPG) based on a ppLN waveguide. The idler spectrum is measured using a dispersive time-of-flight spectrometer in coincidence with a successful projection from the QPG. BPF: Bandpass filter. APD: Avalanche photo-diode. BS: beamsplitter. PBS: polarizing beamsplitter. DM: dichroic mirror. SNSPD: Superconducting nanowire single-photon detectors. OPO: Optical parametric oscillator. SHG: Second harmonic generation.}
\label{fig:expsetup}
\end{figure}

\begin{figure}[t!]
\centering
\includegraphics[width=1\linewidth]{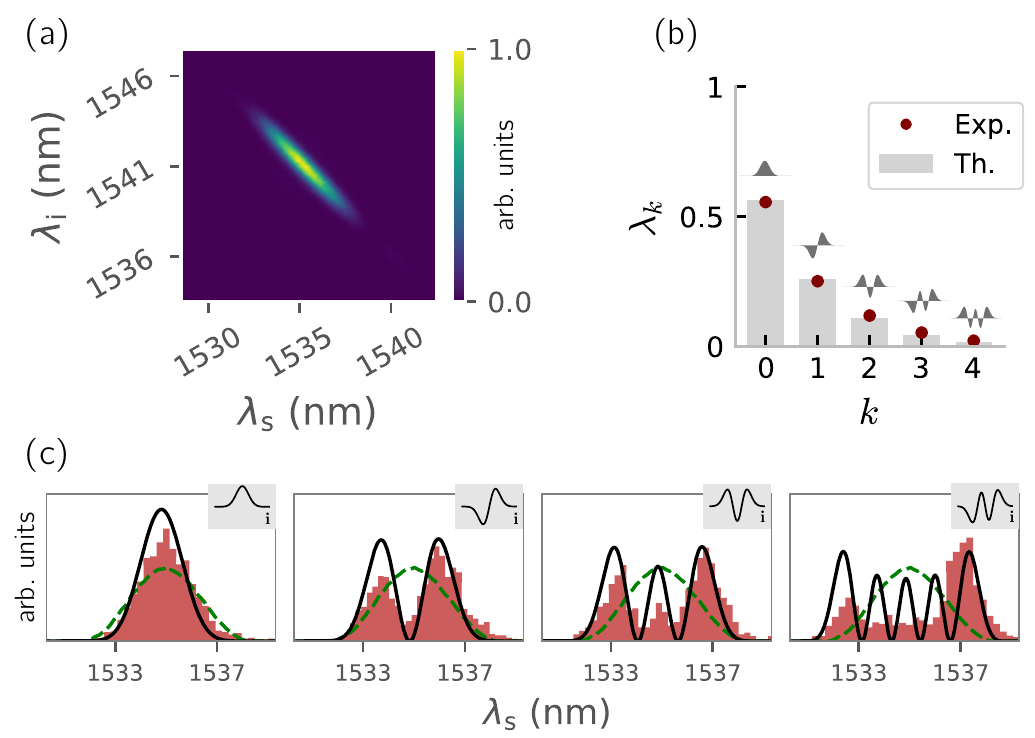}
\caption{(a) The joint spectral intensity of the PDC state with a narrowband pump (Gaussian with standard deviation of 0.3~nm), displaying the expected frequency anti-correlations. (b) Tomographically reconstructed Schmidt mode decomposition of the idler photon revealing the underlying temporal modes. Error bars are smaller than the plotted points. (c) Conditional spectra of signal photon when the idler is projected onto the first four Hermite-Gauss temporal modes (indicated in insets) is shown in red, closely matching the theoretical expectation (black line). As a comparison, in dashed green lines we plot the measured marginal spectrum of the signal photons. All conditioned spectra are plotted from approximately 1500 detection events.}
\label{fig:pdc0_}
\end{figure}

\begin{figure}[t!]
\centering
\includegraphics[width=1\linewidth]{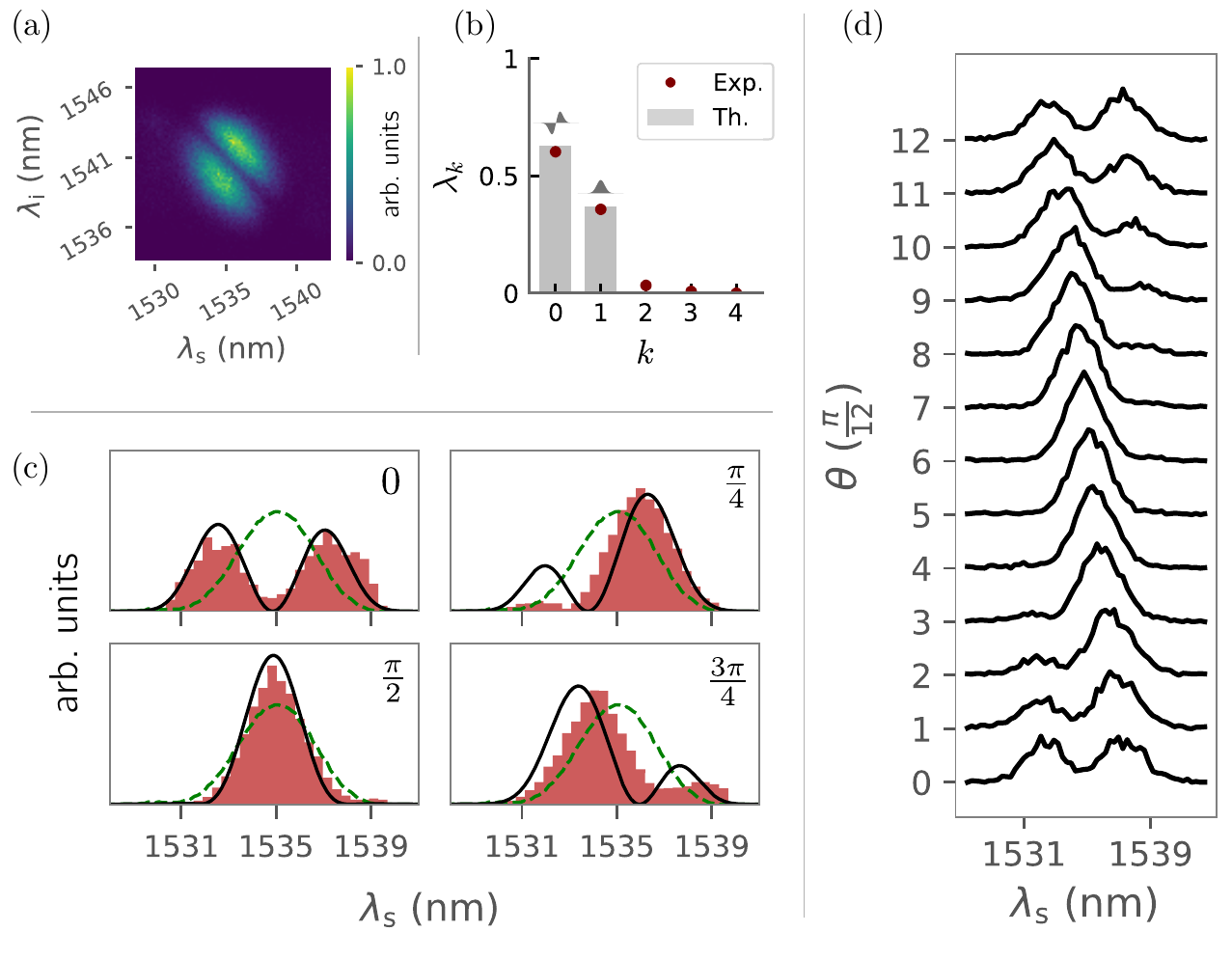}
\caption{Remote state preparation from a time-frequency Bell state. (a) The PDC joint spectral intensity, and (b) tomographically reconstructed Schmidt mode decomposition of the idler photon, with error bars smaller than the plotted points. Notably, the tomography shows that mainly two time-frequency modes (the zeroth and first order Hermite-Gauss) are present. (c) The signal spectrum (in red) conditioned on projections of the form $\cos\theta\ket{\protect\zeroOrder_\mathrm{i}}+\sin\theta\ket{\protect\firstOrder_\mathrm{i}}$, for $\theta=\left\{0,\frac{\pi}{4},\frac{\pi}{2},\frac{3\pi}{4}\right\}$, respectively. The theoretical expectation is given by the black line, and measured non-mode-resolved heralded spectrum of the signal photons in dashed green lines. (d) While varying the superposition weight $\theta$, a continuous shift in the conditional signal spectrum is observed. The conditioned spectra in (c,d) are extracted from between 2900 and 5800 detection events, depending on the projection mode.}
\label{fig:result}
\end{figure}

To generate time-frequency entangled states, we use a guided-wave PDC source; an 8 mm long periodically poled potassium titanyl phosphate waveguide (ppKTP; poling period 117~$\mathrm{\mu}$m; room temperature; spatially single-mode for PDC photons), see Fig.~\ref{fig:expsetup}. This versatile source allows us to create PDC states with different forms of time-frequency correlations by only reshaping pump pulses, as we have shown previously \cite{Eckstein2011b, Harder2013, ansari2018tomography}. Orthogonally polarized down-converted PDC photons, signal and idler, are separated using a polarizing beamsplitter and subsequently coupled into single-mode fibers and time-of-flight spectrometers \cite{Avenhaus2009a}, comprised of highly dispersive fibers followed by superconducting nanowire single-photon detectors (SNSPD, PhotonSpot). Here, we consider two different PDC states (see Fig.~\ref{fig:schmidt}): (A) with traditional frequency anti-correlations and (B) a pulsed temporal-mode Bell-like state. The joint spectral intensity (JSI) of these states, measured by time-of-flight spectrometers, are plotted in Fig.~\ref{fig:pdc0_}(a) and Fig.~\ref{fig:result}(a), for states (A) and (B), respectively. The JSI spectra closely resemble the expected theory, showing that the system behaves as expected. Note that a full characterization would require a measurement of the JSA, which would include spectral phase information. 

To realize temporal-mode selection, in the following, we couple the idler photon to the QPG, as shown in Fig.~\ref{fig:expsetup}. The QPG, as discussed in the previous section, is a time-frequency mode-selective sum-frequency generation process. Our QPG implementation, shown in Fig.~\ref{fig:expsetup}, comprises a 17 mm periodically poled lithium niobate waveguide (ppLN; spatially single-mode at 1540 nm; poling period 4.4~$\mathrm{\mu}$m; temperature 470~K) with pump pulses synchronized to the input field i.e. PDC idler photons. The pulsed temporal-mode that the QPG selects is controlled by the shape of its pump pulse, see Eq.~\eqref{eq:proj}. To shape the spectral phase and amplitude of pump pulses, we use a diffractive spatial light modulator (Hamamatsu LCoS) at the focal plane of a folded 4f setup. At the output port of the QPG, the sum-frequency pulse at 558~nm is isolated using a 4f filter and coupled into a single-mode fiber, where it is detected with a silicon avalanche photo-diode (SiAPD, ID Quantique ID100). 

To characterize the underlying pulsed temporal-modes of the generated PDC states, we use a previously developed tomographic method based on the QPG \cite{ansari2018tomography, ansari2017temporal}. Performing an informationally complete set of measurements on the idler photons we can reconstruct its reduced density matrix $\rho_{\mathrm{i}}=\mathrm{tr}_{\mathrm{s}}[\rho_{\mathrm{PDC}}]$. The eigenvalues of the idler photon's reduced density matrices for states (A) and (B) are plotted in Fig.~\ref{fig:pdc0_}(b) and Fig.~\ref{fig:result}(b), respectively. Through these measurements we determine the idler's modal decomposition, with its fundamental Gauss-mode centered at 1540.7~nm with a standard deviation of 0.84~nm and 1.43~nm for states (A) and (B), respectively.

The key to our experiment is the use of the QPG to project one PDC photon into a chosen Schmidt mode composition. Heralded on such projective measurements, we record the spectrum of the PDC's partner photon, as shown in Fig.~\ref{fig:outline}. Such conditional spectral measurements on the signal photons is achieved by collecting coincident detection events between the time-of-flight spectrometer (detecting the idler photon) and the SiAPD at the output of QPG. Our time-of-flight spectrometer converts a spectral shift of 1~nm to a time delay of 0.58~ns. The overall theoretical-resolution of our experiment is 0.15~nm, limited by the timing jitter of the SNSPDs, as well as the jitter of the triggering pulses from the SiAPD. Another source of timing uncertainties is the mechanical drifts and instabilities of the setup, which can jitter the timing between between the QPG pump pulse and the PDC photon. To minimize such errors, we take our measurements over relatively short time scales in which the setup is adequately stable.

\section{Results and discussion}
In this section we present remote shaping of PDC photons, considering the states (A) and (B) illustrated in Fig.~\ref{fig:schmidt}. Our experimental result for the traditional two-photon state with frequency anti-correlations, state (A), is shown in Fig.~\ref{fig:pdc0_}(c). When we project onto one of the Hermite-Gauss Schmidt modes of the idler photon, we find that the measured spectrum of the signal photon is as expected for the corresponding Schmidt mode. This correspondence is clear for the first three modes, but weakens for the fourth due to low count rates arising from the small Schmidt coefficient. 

To show that we can remotely prepare superpositions of spectral amplitudes, in the following, we use state (B) with pulsed temporal-mode Bell-like correlation. The reduced state of the idler photon is almost completely described as a mixture of the zeroth- and first-order Hermite-Gauss modes. Note that, in the experiment, the relative amplitude between the two modes is unbalanced due to the imperfect symmetric group-velocity matching of the ppKTP source Fig.~\ref{fig:schmidt}. To remotely shape the signal photon's spectrum, we project the idler photon into superposition states of the Hermite-Gauss modes. We start our experiments by operating on the computational basis of our maximally entangled temporal-mode state. As seen in Fig.~\ref{fig:result}(c), when we project the idler into a superposition state, the signal spectrum takes the form of the conjugate superposition, consistent to our time-frequency entangled state. As a comparison, we also plot the measured marginal spectrum of the signal photon in dashed green lines, corresponding to a non-mode-resolved measurement. Furthermore, in Fig.~\ref{fig:result}(d), we project upon twelve superposition states, spanning the x-z plane of the Bloch sphere, and show a continuous reshaping of the signal spectrum dependent on the projection employed.

\begin{figure}[t!]
\centering
\includegraphics[width=1\linewidth]{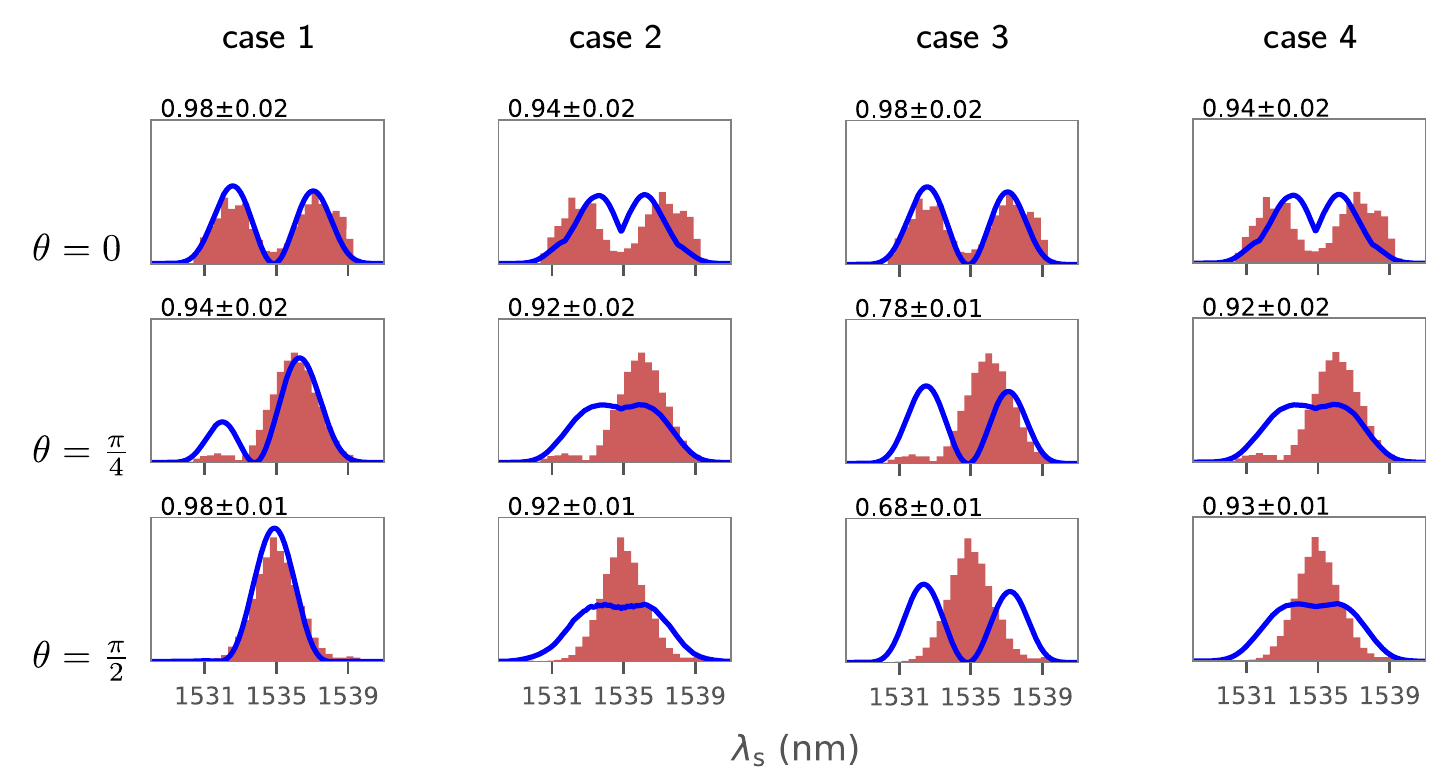}
\caption{Comparison of our experimental results with four theory models: coherent, partially coherent, and incoherent multi-mode theories. The signal spectrum (in red) is conditioned on idler projections with $\theta=0$ (top row), $\theta=\frac{\pi}{4}$ (middle row), and $\theta=\frac{\pi}{2}$ (bottom row). The solid blue lines are modeled by four different theory models detailed in the main text. To quantify the level of agreement between the experimental outcome and different models, we print the similarity value next to each plot, calculated as $| \int d\lambda_\mathrm{s} \sqrt{ f(\lambda_\mathrm{s}) \tilde{f}(\lambda_\mathrm{s}) } |$.}
\label{fig:coherent_v_incoherent}
\end{figure}

Both measurements described above show that the system behaves as expected, and that we can remotely prepare photons with any modal composition. While these measurement results are in excellent agreement with the coherent multi-mode theory, they lack phase information and assume coherence in the state generation and the mode-selective projections.

The coherence of the system can be probed without resorting to a full JSA analysis. One can test the presence of coherence by assuming that coherence is not present in multiple ways and compare the measured results. In the following, we compare our experimental results with four possible theoretical settings:
\begin{itemize}
	\item case 1: the two-photon state is the entangled state (B), and we project upon temporal modes coherently, as formulated throughout the previous sections; 
	\item case 2: the two-photon state is an incoherent mixture of the Schmidt modes of the entangled state (B), and we project upon temporal modes coherently; 
	\item case 3: the two-photon state is the entangled state (B), but our measurement is intensity-only spectral filtering with the same shape as the spectral intensity of the intended temporal modes (i.e. incoherent measurements); 
	\item case 4: the two-photon state is an incoherent mixture, and the measurements are incoherent intensity-only measurements.
\end{itemize}
In Fig.~\ref{fig:coherent_v_incoherent}, we plot our experimental results versus the simulated outcomes from the above four cases. Although among all cases for specific projections we can find resemblance between theory and experiment, only the fully coherent model (case 1) shows a consistent agreement with our experimental data. For a more qualitative comparison, we print the similarity value between theory and experiment, above each plot. This shows the essential role of coherence in state preparation and mode-resolved measurements.

\section{Conclusion}
We have shown time-frequency reshaping of PDC photons through mode-selective measurements. This technique allows remote shaping of PDC photons into any coherent composition of their Schmidt modes. This can be further extended with highly multi-mode and entangled photon pair sources for quantum communication protocols and more complex remote shaping of single-photon states in quantum networks.

\section*{Funding}
This research has received funding from the Gottfried Wilhelm Leibniz-Preis, the Natural Sciences and Engineering Research Council of Canada (NSERC), and from the European Union's Horizon 2020 research and innovation programme under grant agreement No 665148. 

\section*{Acknowledgments}
We acknowledge helpful discussions with Jan Sperling, Alexander Streltsov, and Michael Stefszky.


%

\end{document}